# Correlation energy of many-electron systems: a modified Colle-Salvetti approach

Sébastien RAGOT and Pietro CORTONA

Laboratoire Structure, Propriété et Modélisation des Solides (CNRS, Unité Mixte de Recherche 85-80). École Centrale Paris, Grande Voie des Vignes, 92295 CHATENAY-MALABRY, FRANCE

## Abstract

The Colle and Salvetti approach [Theoret. Chim. Acta, **37**, 329 (1975)] to the calculation of the correlation energy of a system is modified in order to explicitly include into the theory the kinetic contribution to the correlation energy. This is achieved by deducing from a many electrons wave function, including the correlation effects via a Jastrow factor, an approximate expression of the one-electron reduced density matrix. Applying the latter to the homogeneous electron gas, an analytic expression of the correlation kinetic energy is derived. The total correlation energy of such a system is then deduced from its kinetic contribution inverting a standard procedure. At variance of the original Colle-Salvetti theory, the parameters entering in both the kinetic correlation and the total correlation energies are determined analytically, leading to a satisfactory agreement with the results of Perdew and Wang [Phys. Rev. B **45**, 13244 (1992)]. The resulting (parameter-free) expressions give rise to a modified-local-density approximation that can be used in self-consistent density-functional calculations. We have performed such calculations for a large set of atoms and ions and we have found results for the correlation energies and for the ionization potentials which improve those of the standard local-density approximation.

**Keywords**: Density matrices, correlation energy functional, kinetic energy, uniform electron gas.

# Introduction

The basic idea of the Colle-Salvetti (CS) approach for the calculation of the correlation energy of many-electron systems is the following: from a correlated wave function, an approximate two-electron reduced density matrix (2-RDM) is derived, which allows the computation of the correlation energy [1]. The resulting 2-RDM takes the form of an uncorrelated 2-RDM multiplied by a correlation factor [2]. As the uncorrelated 2-RDM is taken to be the Hartree-Fock (HF) one, the CS correlation energy becomes a functional of the HF one-electron reduced density matrix (1-RDM). It is further assumed that the correlation effects on the 1-RDM can be neglected: as a consequence, the correlation kinetic energy is assumed to be zero. The only contribution to the correlation energy comes from the electron-electron potential energy, computed from the model 2-RDM. Nevertheless, the CS approach allows accurate calculations of the correlation energies of atoms and small molecules. This is achieved by a suitable determination of the only parameter involved in the final CS expression. This parameter, generally indicated by $q$, is related to the size of the Coulomb-correlation hole.

Because of both its simplicity and accuracy, the CS approach has received considerable attention [3,4]. However, it has been shown [5-7] that it suffers from physical inconsistencies, many of which result from an incorrect normalization of the correlated 2-RDM.

As pointed out above, the CS approach neglects the kinetic part of the correlation energy. In contrast, the analysis of the total correlation energy ($E_c$) of first and second-row atoms reveals two dominant contributions [8], which arise from the correlation corrections to the HF electron-pair potential energy ($V_c^{ee}$) and to the HF kinetic energy ($T_c$). For such systems, the correlation correction $T_c$ is approximately half the magnitude of $V_c^{ee}$. Thus, imposing any parameterized form for $V_c^{ee}$ to equal $E_c$ amounts to underestimate the true $V_c^{ee}$ by a factor of about 2. Such shortcomings are of little importance as long as one is interested in estimating the whole correlation energy. However, as outlined by Singh and co-workers [5]: "*Although the CS correlation energies are often accurate and, in that sense pragmatic; nonetheless, there remains the deeper question of whether or not the physics of the separate components of the correlation energy are described correctly*".

The aim of this work is to show that the kinetic correlation energy can be accounted for in a CS-like approach. Of course, many alternative ways of accessing the kinetic energy of correlation exists (see for example [9,10]). The present work focuses on the 1-RDM: one of our purposes is to improve the understanding of the correlation effects on it.

The paper is organized as follows. After recalling some definitions (sect. I), we derive an approximate expression for the correlated 1-RDM (sect. II). This approximate 1-RDM is then applied to the UEG and analytic expressions of the kinetic and the total correlation energies are worked out (sect. III). In the last section, we use the formulae derived for the UEG in order to perform self-consistent calculations of correlation energies and ionization potentials of several atoms. Atomic units are used throughout.

## I. Definitions

Let $x_i$ denote the space and spin coordinates of the electron $i$. The 1- and 2-RDMs derived from a many-electron wave function $\psi$ are defined by [11,12]:

$$\gamma_1(x_1; x_1') = N \int \psi(x_1, x_2, ..., x_N) \psi^*(x_1', x_2, ..., x_N) dx_2 ... dx_N \tag{1}$$

and

$$\gamma_2(x_1, x_2; x_1', x_2') = \frac{N(N-1)}{2} \int \psi(x_1, x_2, x_3, ..., x_N) \psi^*(x_1', x_2', x_3, ..., x_N) dx_3 ... dx_N \tag{2}$$

respectively. Integrating (1) and (2) over the spin variables leads to the spinless RDMs:

$$\rho_1(\mathbf{r}_1; \mathbf{r}_1') = \int [\gamma(x_1; x_1')]_{s_1=s_1'} ds_1 \tag{3}$$

$$\rho_2(\mathbf{r}_1, \mathbf{r}_2; \mathbf{r}_1', \mathbf{r}_2') = \int [\gamma(x_1, x_2; x_1', x_2')]_{s_1=s_1', s_2=s_2'} ds_1 ds_2 \tag{4}$$

For simplicity, the short notation «RDMs» will hereafter apply for the spinless matrices as well. The diagonal elements of expressions (3) and (4) are the electron charge and pair densities, respectively denoted by:

$$\rho(\mathbf{r}_1) = \rho_1(\mathbf{r}_1; \mathbf{r}_1) \tag{5}$$

$$P(\mathbf{r}_1, \mathbf{r}_2) = \rho_2(\mathbf{r}_1, \mathbf{r}_2; \mathbf{r}_1, \mathbf{r}_2) \tag{6}$$

Note that the definitions (1) and (2) imply the condition:

$$\rho_1(\mathbf{r}_1; \mathbf{r}_1') = \frac{2}{N-1} \int \rho_2(\mathbf{r}_1, \mathbf{r}_2; \mathbf{r}_1', \mathbf{r}_2) d\mathbf{r}_2 \tag{7}$$

The electron-electron potential energy is given by:

$$V^{ee} = \int \frac{P(\mathbf{r}_1, \mathbf{r}_2)}{r_{12}} d\mathbf{r}_1 d\mathbf{r}_2 \tag{8}$$

where $r_{12} = |\mathbf{r}_1 - \mathbf{r}_2|$.

The RDMs derived from a single determinant can be expressed from the 1-RDM only [12]. In particular:

$$\gamma_2^{HF}(x_1,x_2;x_1',x_2') = \frac{1}{2}\left[\gamma_1^{HF}(x_1;x_1')\gamma_1^{HF}(x_2;x_2') - \gamma_1^{HF}(x_1;x_2')\gamma_1^{HF}(x_2;x_1')\right] \quad (9)$$

Thus, the HF pair density obtained from eq. (9) reads:

$$P^{HF}(\mathbf{r}_1,\mathbf{r}_2) = \frac{1}{2}\left[\rho_1^{HF}(\mathbf{r}_1,\mathbf{r}_1)\rho_1^{HF}(\mathbf{r}_2,\mathbf{r}_2) - \frac{1}{2}\rho_1^{HF}(\mathbf{r}_1,\mathbf{r}_2)\rho_1^{HF}(\mathbf{r}_2,\mathbf{r}_1)\right] \quad (10)$$

In the following, we shall call a "correlation quantity" the difference between the exact non-relativistic quantity and the HF-approximated one. The electron-electron potential energy of correlation is thus:

$$V_c^{ee} = \int \frac{P(\mathbf{r}_1,\mathbf{r}_2) - P^{HF}(\mathbf{r}_1,\mathbf{r}_2)}{r_{12}} d\mathbf{r}_1 d\mathbf{r}_2 \quad (11)$$

We now turn to the kinetic energy of correlation, the definition of which is natural in momentum space. First, we rewrite $\rho_1(\mathbf{r}_1;\mathbf{r}_1')$ in terms of center-of-mass and relative coordinates, *i.e.* in its intracular-extracular representation [13]:

$$\rho_1(\mathbf{r}_1;\mathbf{r}_1') = \tilde{\rho}_1(\mathbf{R},\mathbf{s}) \quad (12)$$

where **R** stands for (**r** + **r'**)/2 and **s** is the difference **r** - **r'**. The momentum density is defined as

$$n(\mathbf{p}) = \frac{1}{(2\pi)^3}\int \rho_1(\mathbf{r};\mathbf{r}')e^{i\mathbf{p}\cdot(\mathbf{r}-\mathbf{r}')}d\mathbf{r}d\mathbf{r}' = \frac{1}{(2\pi)^3}\int \tilde{\rho}_1(\mathbf{R},\mathbf{s})e^{i\mathbf{p}\cdot\mathbf{s}}d\mathbf{R}d\mathbf{s} \quad (13)$$

Thus, $n(\mathbf{p})$ is the Fourier transform of the so-called auto correlation function [14,15], which is obtained by integrating $\tilde{\rho}_1(\mathbf{R},\mathbf{s})$ with respect to the extracular coordinate:

$$B(\mathbf{s}) = \int \tilde{\rho}_1(\mathbf{R},\mathbf{s})d\mathbf{R} \quad (14)$$

so that, from eq. (13):

$$B(\mathbf{s}) = \int n(\mathbf{p})e^{-i\mathbf{p}\cdot\mathbf{s}}d\mathbf{p} \quad (15)$$

Note that from eqs. (5) and (12), we can derive a simple relation between the electron charge density and the intracular-extracular representation of the spinless 1-RDM: $\rho(\mathbf{R}) \equiv \tilde{\rho}_1(\mathbf{R},\mathbf{0})$. This relation implies the following normalization condition:

$$B(\mathbf{0}) = \int \tilde{\rho}_1(\mathbf{R},\mathbf{0})d\mathbf{R} = \int \rho(\mathbf{R})d\mathbf{R} = N \quad (16)$$

or conversely, from eq. (15):

$$B(\mathbf{0}) = \int n(\mathbf{p})d\mathbf{p} = N \quad (17)$$

Thus, the condition $B(\mathbf{0}) = N$ ensures a correct normalization of the one-electron distributions. Next, using eq. (15), one can verify that:

$$-\frac{1}{2}\nabla_s^2 B(\mathbf{s})\Big|_{\mathbf{s}=\mathbf{0}} = \int \frac{p^2}{2} n(\mathbf{p}) d\mathbf{p} = T \tag{18}$$

which shows that the kinetic energy is entirely determined by the behavior of the auto-correlation function near $\mathbf{s} = \mathbf{0}$ [16]. The kinetic energy of correlation is then given by:

$$T_c = \int \frac{p^2}{2} \left[ n(\mathbf{p}) - n^{HF}(\mathbf{p}) \right] d\mathbf{p} = -\frac{1}{2}\nabla_s^2 B(\mathbf{s})\Big|_{\mathbf{s}=\mathbf{0}} - \frac{1}{2}\nabla_s^2 B^{HF}(\mathbf{s})\Big|_{\mathbf{s}=\mathbf{0}} \tag{19}$$

Moreover, for spherically symmetric systems, eq. (18) reduces to:

$$T = -\frac{3}{2} B''(0) \tag{20}$$

This simple expression for the kinetic energy of a spherically symmetric system is a key element in the present contribution. In the following, we derive a simple approximation for the auto-correlation function of a correlated uniform electron gas near $s = 0$, which, in turn, shall allow us to derive an analytical expression for the kinetic energy through eq. (20). To achieve this aim, we first need an expression for the correlated 1-RDM.

## II. Correlated reduced density matrices

The CS correlated ground-state wave function is written as [1,17]:

$$\psi(x_1, \ldots, x_N) = \psi^{HF}(x_1, \ldots, x_N) \prod_{i<j} \left[ 1 - f(\mathbf{r}_i, \mathbf{r}_j) \right] \tag{21}$$

where $f(\mathbf{r}_i, \mathbf{r}_j)$ correlates the electron pair $(i, j)$ regardless of the spin coordinates of both electrons, and $\psi^{HF}$ denotes the HF approximation to the ground-state wave function. Assuming real functions $f(\mathbf{r}_i, \mathbf{r}_j)$, it is straightforward to show that $\rho_2$ can be written as follows:

$$\rho_2(\mathbf{r}_1, \mathbf{r}_2; \mathbf{r}_1', \mathbf{r}_2') \tag{22}$$
$$= \rho_2^{HF}(\mathbf{r}_1, \mathbf{r}_2; \mathbf{r}_1', \mathbf{r}_2') \left[ 1 - f(\mathbf{r}_1, \mathbf{r}_2) - f(\mathbf{r}_1', \mathbf{r}_2') + f(\mathbf{r}_1, \mathbf{r}_2) f(\mathbf{r}_1', \mathbf{r}_2') \right] + R(\mathbf{r}_1, \mathbf{r}_2; \mathbf{r}_1', \mathbf{r}_2')$$

where $\rho_2^{HF}$ is the HF 2-RDM and $R$ includes all the terms for which $\rho_2^{HF}$ can not be factorized out [18]. The CS model takes a truncated form of eq. (22):

$$\rho_2(\mathbf{r}_1, \mathbf{r}_2; \mathbf{r}_1', \mathbf{r}_2') = \rho_2^{HF}(\mathbf{r}_1, \mathbf{r}_2; \mathbf{r}_1', \mathbf{r}_2') \left[ 1 - f(\mathbf{r}_1, \mathbf{r}_2) - f(\mathbf{r}_1', \mathbf{r}_2') + f(\mathbf{r}_1, \mathbf{r}_2) f(\mathbf{r}_1', \mathbf{r}_2') \right] \tag{23}$$

as a starting point for the derivation of a correlation energy expression. Eq. (23) is exact for 2-electron systems (to a normalization factor) but, for larger systems, it neglects the $N$-electron effects on the 2-RDM beyond the direct pair interactions. In this sense, the CS approach has a close connection with the independent pair approximation (IPA), which is known to be correct to first order in the correlation function $f$ [19]. Moreover, it has been shown that the 2-RDM, eq. (23), is not $N$-representable [20]. An important consequence of this is that the 1-RDM derived from eq. (7) and eq. (23) differs from that obtained by putting in eq. (1) the wave function defined in eq. (21). In fact, in the first case one has:

$$\rho_1(\mathbf{r}_1;\mathbf{r}_1') = \rho_1^{HF}(\mathbf{r}_1;\mathbf{r}_1') \qquad (24)$$
$$+ \frac{2}{N-1}\int [-f(\mathbf{r}_1,\mathbf{r}_2) - f(\mathbf{r}_1',\mathbf{r}_2) + f(\mathbf{r}_1,\mathbf{r}_2)f(\mathbf{r}_1',\mathbf{r}_2)]\rho_2^{HF}(\mathbf{r}_1,\mathbf{r}_2;\mathbf{r}_1',\mathbf{r}_2)d\mathbf{r}_2$$

while in the second one the following expression is obtained [18]:

$$\rho_1(\mathbf{r}_1;\mathbf{r}_1') = \rho_1^{HF}(\mathbf{r}_1;\mathbf{r}_1') \qquad (25)$$
$$+ 2\int [-f(\mathbf{r}_1,\mathbf{r}_2) - f(\mathbf{r}_1',\mathbf{r}_2) + f(\mathbf{r}_1,\mathbf{r}_2)f(\mathbf{r}_1',\mathbf{r}_2)]\rho_2^{HF}(\mathbf{r}_1,\mathbf{r}_2;\mathbf{r}_1',\mathbf{r}_2)d\mathbf{r}_2 + ...$$

The comparison of expressions (24) and (25) shows that the CS approximation to the 2-RDM is not an appropriate starting point for the calculation of the correlated 1-RDM: it widely underestimates the correlation effects on the 1-RDM and so, the kinetic part of the correlation.

It should be noted that neither the 2-RDMs, eqs. (22) and (23), nor the wave function given in eq. (21) are normalized. Thus, the expression that we finally retain for the model correlated 1-RDM is:

$$\rho_1(\mathbf{r}_1;\mathbf{r}_1') = \mathcal{N}\{\rho_1^{HF}(\mathbf{r}_1;\mathbf{r}_1')$$
$$+ 2\int [-f(\mathbf{r}_1,\mathbf{r}_2) - f(\mathbf{r}_1',\mathbf{r}_2) + f(\mathbf{r}_1,\mathbf{r}_2)f(\mathbf{r}_1',\mathbf{r}_2)]\rho_2^{HF}(\mathbf{r}_1,\mathbf{r}_2;\mathbf{r}_1',\mathbf{r}_2)d\mathbf{r}_2\} \qquad (26)$$

where $\mathcal{N}$ is a normalization factor. This approximation of the 1-RDM includes some electron correlation effects beyond the IPA and should provide a reasonable approximation for a uniform electron gas, at least when the density is not too high (see the discussion in the appendix).

# III. Modeling momentum-space properties of the uniform electron gas

### III.1 Approximate 1-matrix

Consider the case of a uniform electron gas (UEG), whose constant density $\rho$ can be related either to the Seitz radius $r_s$ or to the Fermi momentum $k_F$:

$$\rho = \frac{3}{4\pi r_s^3} = \frac{k_F^3}{3\pi^2} \tag{27}$$

The spinless 1-RDM solely depends on $s = |\mathbf{r} - \mathbf{r}'|$ and, from eq. (14), is proportional to $B(s)$. A convenient normalization choice for the UEG is to set $B(0) = 1$. Using periodic boundary conditions, the HF description of the UEG yields to [21]:

$$B^{HF}(s) = 3\frac{\sin(k_F s) - k_F s \cos(k_F s)}{(k_F s)^3} \tag{28}$$

Such a function depends implicitly on $\rho$ through $k_F$, has a maximum at $s = 0$, and vanishes as $s \to \infty$, while oscillating (see fig. 1). Now, the HF pair density of the UEG can be explicitly stated as:

$$P^{HF}(r_{12}) = \frac{1}{2}\rho^2\left\{1 - \frac{1}{2}[B^{HF}(r_{12})]^2\right\} = \frac{1}{2}\rho^2 g^{HF}(r_{12}) \tag{29}$$

The term in the curly brackets ($g^{HF}$) is often referred to as the HF pair-distribution function. It satisfies the normalization condition:

$$\int \rho[g^{HF}(r_{12}) - 1]d\mathbf{r}_{12} = -1 \tag{30}$$

The HF approximation generates the so-called "exchange hole" around the position of any reference electrons. This kind of correlation arises from the determinantal nature of the wave function but does not include the correlation beyond the Pauli level.

The next step consists of including the Coulomb correlation by means of a Jastrow-like correlation factor, which Colle and Salvetti formulated as:

$$f(\mathbf{r}_1, \mathbf{r}_2) = \left[1 - \Phi(\mathbf{R}_{12})\left(1 + \frac{r_{12}}{2}\right)\right] e^{-\beta_c^2(\mathbf{R}_{12})r_{12}^2} \tag{31}$$

Such a correlation function has a cusp in $r_{12} = 0$ and rapidly falls off as $r_{12}$ increases, because of the gaussian term. In eq. (31), $\mathbf{R}_{12}$ is the pair center-of-mass vector $\mathbf{R}_{12} = (\mathbf{r}_1 + \mathbf{r}_2)/2$ and $\beta_c$ has the meaning of an inverse radius of the correlation hole. Colle and Salvetti assumed this radius to be proportional to $r_s$, which amounts to assume $\beta_c(\mathbf{R}_{12}) = q[\rho(\mathbf{R}_{12})]^{1/3}$, where $q$ is to be parameterized. In the case of a UEG, eq. (31) can be rewritten as:

$$f(r_{12}) = \left[1 - \Phi\left(1 + \frac{r_{12}}{2}\right)\right] e^{-\beta_c^2 r_{12}^2} \tag{32}$$

where $\beta_c$ and $\Phi$ do not depend on $\mathbf{R}_{12}$ but still depend on $\rho$ or, equivalently, on $r_s$. In order to determine $\Phi$, Colle and Salvetti assumed that $\rho_1(\mathbf{r}_1,\mathbf{r}_1) = \rho_1^{HF}(\mathbf{r}_1,\mathbf{r}_1)$. This condition, which is exact for the UEG, but not for atomic systems [5], is verified if the correction term to $\rho_1^{HF}(\mathbf{r}_1,\mathbf{r}_1)$ in eq. (24) does cancel, that is, if:

$$\int [-f(\mathbf{r}_1,\mathbf{r}_2) - f(\mathbf{r}_1,\mathbf{r}_2) + f(\mathbf{r}_1,\mathbf{r}_2)f(\mathbf{r}_1,\mathbf{r}_2)]\rho_2^{HF}(\mathbf{r}_1,\mathbf{r}_2;\mathbf{r}_1,\mathbf{r}_2)d\mathbf{r}_2 = 0 \tag{33}$$

Using further approximations, they finally obtained:

$$\Phi = \frac{\sqrt{\pi}\beta_c}{(1 + \sqrt{\pi}\beta_c)} \tag{34}$$

This expression ensures the correct asymptotic limits that we can expect for $\Phi$, which varies between 0 and 1 as $\beta_c$ (and therefore $\rho$) goes from 0 to infinity.

Since (34) is not the exact solution to (33), it is interesting to understand its exact physical meaning. Consider, for instance, in eq. (24), the first-order correction term (in $f$) to the Hartree-Fock 1-RDM:

$$\int [f(\mathbf{r}_1,\mathbf{r}_2) + f(\mathbf{r}_1',\mathbf{r}_2)]\rho_2^{HF}(\mathbf{r}_1,\mathbf{r}_2;\mathbf{r}_1',\mathbf{r}_2)d\mathbf{r}_2 \tag{35}$$

The term $\rho_2^{HF}(\mathbf{r}_1,\mathbf{r}_2;\mathbf{r}_1',\mathbf{r}_2)$ separates into two components:

$$\rho_2^{HF}(\mathbf{r}_1,\mathbf{r}_2;\mathbf{r}_1',\mathbf{r}_2) = \frac{1}{2}\rho^2\left[B^{HF}(r_{11'}) - \frac{1}{2}B^{HF}(r_{12})B^{HF}(r_{1'2})\right] \tag{36}$$

The first-order correlation correction term, given in eq. (35), can thus be rewritten as a sum of two terms: a first-order Coulomb-correlation term:

$$\frac{1}{2}\rho^2 B^{HF}(r_{11'})\int [f(r_{12}) + f(r_{1'2})]d\mathbf{r}_2 \tag{37}$$

and an exchange-correlation term:

$$-\frac{1}{4}\rho^2 \int [f(r_{12}) + f(r_{1'2})]B^{HF}(r_{12})B^{HF}(r_{1'2})d\mathbf{r}_2 \tag{38}$$

Imposing the first-order Coulomb-correlation term (37) to cancel yields to the condition:

$$\int f(r_{12})d\mathbf{r}_2 = \int f(r_{1'2})d\mathbf{r}_2 = 0 \tag{39}$$

Using (32), it can be verified that the condition (39) is satisfied if $\Phi = \sqrt{\pi}\beta_c/(1+\sqrt{\pi}\beta_c)$: this result is identical to that of CS. Thus, expression (34) implies that the first-order Coulomb-correlation correction to the 1-RDM must cancel. For practical reasons, we shall make use of the expression (34) in the following.

Incidentally, the CS approach is completed by assuming that:

$$\rho_1(\mathbf{r}_1;\mathbf{r}_1') = \rho_1^{HF}(\mathbf{r}_1;\mathbf{r}_1') \tag{40}$$

Such a condition might, at first, seem reasonable, since correlation effects on $\rho_1(\mathbf{r}_1;\mathbf{r}_1')$ are known to be of "second order" while being of "first order" on the pair density [19,22]. However, as discussed above, condition (40) implies the kinetic correlation energy to be zero and, thus, makes $V_c^{ee}$ the only contribution to the correlation energy.

The damping by the gaussian function in eq. (32) suggests that a gaussian approximation may be made to $B^{HF}(s)$, in the spirit of refs. [2, 23-25]. As we shall see now, the gaussian approximation simplifies considerably the derivation of an analytical expression for the correlation kinetic energy, while ensuring a correct normalization of $n^{HF}(p)$ together with the exact value of the HF kinetic energy. For small $s$, $B^{HF}(s)$ behaves like $B^{HF}(s) \approx 1 - \beta_x^2 s^2$, with $\beta_x = C_x \rho^{1/3}$ and $C_x = 10^{-1/2}(3\pi^2)^{1/3}$. An appropriate gaussian resummation is thus:

$$B_G^{HF}(s) = e^{-\beta_x^2 s^2} \tag{41}$$

The comparison of $B^{HF}$ and $B_G^{HF}$ is shown in fig. 1. The gaussian approximation bypasses the oscillations of $B^{HF}(s)$ at large $s$. However, since $B^{HF}(s)$ and $B_G^{HF}(s)$ become identical at small $s$, they result in identical kinetic energy expressions, from eq. (20).

Replacing now $B^{HF}(s)$ by $B_G^{HF}(s)$ in eq. (29) leads to the approximate HF pair density:

$$P^{HF}(r_{12}) \approx \frac{1}{2}\rho^2\left\{1 - \frac{1}{2}e^{-2\beta_x^2 r_{12}^2}\right\} \tag{42}$$

Such an approximation gives to $\beta_x^{-1}$ the meaning of an exchange hole radius. Notice that $\beta_c$ is proportional to $\beta_x$:

$$\beta_c = q\rho^{1/3} = q\beta_x/C_x \tag{43}$$

Due to the gaussian approximation, the exchange part of the pair density, given in eq. (42), is not properly normalized, which would require another parameterization of $\beta_x$ [21, 23]. The gaussian approximation in eq. (42) leads to an error of about 5% in the normalization condition (30) of the pair distribution. The choice made here should however be appropriate for the present purpose, since correlation effects on the pair density are expected to be important near $r_{12}=0$ (at least when using the CS correlation functional).

As already mentioned, the calculation of the correlated kinetic energy requires the knowledge of $B(s)$ near $s=0$. From eq. (26), we derive the expression for $B(s)$:

$$B(s) = \mathcal{N}\left[B^{HF}(s) + B^{Corr}(s)\right] \tag{44}$$

where the normalization factor is simply determined by the condition $B(0)=1$ and the correlation correction $B^{Corr}(s)$ can be written as:

$$B^{Corr}(s) = \frac{2}{\rho}\int\left[-f(r_{12})-f(r_{1'2})+f(r_{12})f(r_{1'2})\right]\rho_2^{HF}(\mathbf{r}_1,\mathbf{r}_2;\mathbf{r}_1',\mathbf{r}_2)d\mathbf{r}_2 \tag{45}$$

### III.2 Correlation kinetic energy

Using eq. (41), one can replace $\rho_2^{HF}(\mathbf{r}_1,\mathbf{r}_2;\mathbf{r}_1',\mathbf{r}_2)$ in eq. (45) by:

$$\rho_2^{HF}(\mathbf{r}_1,\mathbf{r}_2;\mathbf{r}_1',\mathbf{r}_2) = \frac{1}{2}\rho^2\left[e^{-\beta_x^2 s^2} - \frac{1}{2}e^{-\beta_x^2\left(2|\mathbf{r}_2-\mathbf{R}|^2 + \frac{1}{2}s^2\right)}\right] \tag{46}$$

where $\mathbf{R}$ and $\mathbf{s}$ have been defined in sect. I. Eq. (46) suggests the introduction of the new variable $\tilde{\mathbf{R}} = \mathbf{r}_2 - \mathbf{R}$. All the quantities appearing in eq. (45) can then be rewritten in terms of the variables $\tilde{\mathbf{R}}$ and $\mathbf{s}$. In order to calculate the kinetic energy, it is sufficient to consider a second order development of the right member of eq. (44) (higher order terms give a vanishing contribution). The integral (with respect to $\tilde{\mathbf{R}}$) is then easily performed by standard mathematical software [26]. The subsequent use of eq. (20) yields to the following result for the kinetic energy per electron:

$$t(r_s,\tilde{q}) = K\frac{\dfrac{B_1}{r_s^2} - \dfrac{B_2(\tilde{q})}{r_s[A(\tilde{q})+20r_s]} - \dfrac{B_3(\tilde{q})}{[A(\tilde{q})+20r_s]^2}}{1 + \dfrac{3}{4\pi}\left\{\dfrac{B_4(\tilde{q})}{[A(\tilde{q})+20r_s]^2}r_s^2 - \dfrac{B_5(\tilde{q})}{[A(\tilde{q})+20r_s]}r_s\right\}} \tag{47}$$

In the expression above, $\tilde{q}=q/C_x$, $K$ and $B_1$ are the following numerical constants:

$$K = \frac{3^{\frac{1}{3}}\pi^{\frac{1}{6}}}{2^{\frac{5}{6}}80} \tag{48a}$$

$$B_1 = 36\sqrt{2\pi} \tag{48b}$$

and $A$ and $B_i$ ($i = 2,\ldots,5$) are functions of $\tilde{q}$ given by:

$$A(\tilde{q}) = 3^{\frac{2}{3}} \sqrt{5} (2\pi)^{\frac{5}{6}} \tilde{q} \tag{48c}$$

$$B_2(\tilde{q}) = \frac{1600\sqrt{5}\left[-3(\tilde{q}^2+2)(\tilde{q}^2+1)+\tilde{q}(3\tilde{q}^2+2)\sqrt{\tilde{q}^2+2}\right]}{(\tilde{q}^2+2)^{7/2}} \tag{48d}$$

$$B_3(\tilde{q}) = \frac{125\sqrt{5}\left[\sqrt{2}(48+7\pi)\tilde{q}^2 - 64\tilde{q}\sqrt{\tilde{q}^2+1}+48\sqrt{2}\right]}{(\tilde{q}^2+1)^{3/2}}$$

$$+ \frac{250\sqrt{10}\left[-48(\tilde{q}^2+2)+32\sqrt{2}(\tilde{q}^2+3)-\pi(7\tilde{q}^2+18)\right]}{\tilde{q}^3} \tag{48e}$$

$$B_4(\tilde{q}) = \frac{250\sqrt{5\pi}}{9}\left\{\frac{\sqrt{2}\left[16(-2+\sqrt{2})+3\pi\sqrt{2}\right]}{\tilde{q}^3} - \frac{\left[16+(16+3\pi)\tilde{q}^2 - 16\sqrt{2}\tilde{q}\sqrt{\tilde{q}^2+1}\right]}{(\tilde{q}^2+1)^{5/2}}\right\} \tag{48f}$$

$$B_5(\tilde{q}) = \frac{800\sqrt{10\pi}\left[-2+\tilde{q}\left(-\tilde{q}+\sqrt{\tilde{q}^2+2}\right)\right]}{9(\tilde{q}^2+2)^{5/2}} \tag{48g}$$

The expression (47) for the correlated kinetic energy is obviously more involved than its HF counterpart, to which it reduces as $q \to \infty$, i.e. as the Coulomb hole volume tends to zero:

$$t\Big|_{q \to \infty} = t^{HF} = 9\frac{\sqrt[3]{\frac{3}{2}}\pi^{\frac{2}{3}}}{20 r_s^2} \tag{49}$$

The expression of the kinetic correlation energy $t_c = t - t^{HF}$ is obtained combining eqs. (47), (48), and (49).

Considering now the two limiting cases, $r_s \to 0$ and $r_s \to \infty$, we have:

$$t_c\Big|_{r_s \to 0} = \frac{D_0(\tilde{q})}{r_s} \tag{50}$$

and

$$t_c\Big|_{r_s \to \infty} = \frac{D_\infty(\tilde{q})}{r_s^2} \tag{51}$$

Both these asymptotic expressions are incorrect, since $t_c\big|_{r_s \to 0}$ should diverge [27,28] like $-\ln r_s$, and $t_c\big|_{r_s \to \infty}$ should decay [29,30] like $r_s^{-3/2}$. This is however not surprising, since the approximation (26) to the 1-RDM is supposed to be valid at intermediate and large $r_s$ only, whereas the CS correlation function is not well-suited for describing the long-range correlation. We can nevertheless expect an intermediate range of densities for which the present approach gives reliable results. The function $D_0(\tilde{q})$ has a simple expression:

$$D_0(\tilde{q}) = \frac{K}{A}\left(\frac{3}{4\pi}B_1 B_5 - B_2\right) = \frac{5(2/3)^{1/3}\left(-6 - 3\tilde{q}^2 + 4\tilde{q}\sqrt{\tilde{q}^2 + 2}\right)}{\pi^{2/3}\tilde{q}(\tilde{q}^2 + 2)^{7/2}} \quad (52)$$

The expected limit, $t_c \to +\infty$ as $r_s \to 0$, imposes the condition $D_0(\tilde{q}) > 0$, which is verified if $\tilde{q} > \tilde{q}^*$, with $\tilde{q}^* = 3\sqrt{2/7}$. In terms of $q$, the critical value $q^*$ (above which the divergence of $t_c$ has the correct sign) is given by $q^* = C_x \tilde{q}^* \approx 1.57$.

Conversely, the expression of $D_\infty(\tilde{q})$ is rather involved. $D_\infty(\tilde{q})$ is a positive function of $\tilde{q}$ ($\tilde{q} > 0$) which has a maximum near $\tilde{q} = 1.15$, that is, near $q \approx 1.12$. Thus, $q$ should be the closer possible to 1.12 in order to maximize $t_c$ at large $r_s$ while it must remain at least equal to $q^*$ for ensuring a correct sign to $t_c$ when $r_s \to 0$. Thus, we shall hereafter consider $q = q^*$ for practical calculations. For such a value, $t_c$ becomes almost equal to:

$$t_c(r_s) = \frac{1}{11.9475 + 14.9062 r_s + 4.8440 r_s^2} \quad (53)$$

The results we have obtained for the kinetic energy of correlation using $q = q^*$ and the original CS value ($q = 2.29$) are compared with the quasi-exact results of Perdew and Wang (PW) [31] in fig. 2. Although the correlation kinetic energies, computed by eq. (47) with $q = q^*$, show a correct behavior, they systematically underestimate the PW values. The recovered kinetic energy of correlation is 95% at $r_s = 0.5$ but falls under 39% for $r_s > 5$ (see table 1). As expected, our approach gives rise to satisfactory results at intermediate densities but fails at both high and low densities.

Provided the model 1-RDM (26) is valid at large $r_s$, the increasing discrepancies with PW's results should be due to the damping by the gaussian function in $f(r_{12})$ rather than to the gaussian approximation used for $B^{HF}(s)$ and subsequently for $\rho_2^{HF}(\mathbf{r}_1, \mathbf{r}_2; \mathbf{r}_1', \mathbf{r}_2')$ in eq. (45). In order to check this, we computed $B^{Corr}(0)$ by using the exact HF pair density or the gaussian-approximated one. The relative differences were found to be less than 5% in the range $0.01 < r_s < 0.1$ and less than 0.6 % for $r_s > 1$; the error then continuously decreases as $r_s$ increases. The fact that the agreement found in fig. 2 becomes poor at large $r_s$ indicates that long-range correlation effects are not properly accounted for in eq. (32), as shown by Tao and co-workers [32].

As a final comment on fig. 2, we notice that the use of values of $q$ greater than the limiting value gives rise to increased discrepancies with respect to the PW results.

### III.3 Total correlation energy

An expression for the total correlation energy per electron $\varepsilon_c$ can be derived from eq. (47). This can be achieved by means of the following relation, valid for the homogeneous electron gas [31]:

$$t_c = -\frac{\partial(r_s \varepsilon_c)}{\partial r_s} \tag{54}$$

Integrating the expression of $t_c$ and dividing by $r_s$, one has:

$$\varepsilon_c(r_s, \tilde{q}) =$$

$$-\frac{K}{24\pi A} \left\{ \frac{3(3B_1 B_5 - 4\pi B_2)r_s \{2\ln r_s - \ln[4\pi A^2 + (160\pi - 3B_5)Ar_s + (3B_4 - 60B_5 + 1600\pi)r_s^2]\}}{r_s^2} \right.$$

$$\left. + \frac{2[3B_1(3B_5^2 - 8\pi B_4) - 4\pi(8\pi B_3 + 3B_2 B_5)]\tan^{-1}\left(\frac{-3AB_5 + 160\pi A + 6B_4 r_s - 120 B_5 r_s + 3200\pi r_s}{A\sqrt{48\pi B_4 - 9B_5^2}}\right)}{r_s\sqrt{\frac{16\pi B_4}{3} - B_5^2}} \right\}$$

$$+\frac{C}{r_s} \tag{55}$$

where $K$, $A$, $B_i$, ($i = 1,...,5$) are the constants and the functions of $\tilde{q}$ defined in the preceding section, and $C$ is the integration constant.

In order to determine the value of $C$, we proceed by analogy to the case of $t_c$. Let us first notice that, for $q = q^*$, $3B_1 B_5 - 4\pi B_2 = 0$. Thus, for this particular value of $q$, the first term in the curly brackets vanishes. The leading term of the remaining expression becomes proportional to $r_s^{-1}$ as $r_s \to 0$, and its sign depends on the value of $C$. In particular, this sign is the correct one if:

$$C < C^* \cong 0.897889 \tag{56}$$

In the case of $t_c$ the best value of $q$ was found to be the limiting value $q^*$. Thus, it seems appropriate to make the analogous choice for $\varepsilon_c$ and to set $C = C^*$. Notice that, for these particular values of $q$ and $C$, $\varepsilon_c$ reduces to:

$$\varepsilon_c = \frac{-0.655868 \tan^{-1}(4.888270 + 3.177037 r_s) + 0.897889}{r_s} \qquad (57)$$

Furthermore, both $t_c$ and $\varepsilon_c$ tend to a finite value as $r_s \to 0$ and we have: $t_c|_{r_s=0} = -\varepsilon_c|_{r_s=0}$.

The results we have obtained by eq. (57) are compared with those of the PW expression in fig. 3. In the same figure, we have also reported the results obtained by using the original CS correlation energy expression:

$$E_c = \int \left\{ [f(\mathbf{r}_1, \mathbf{r}_2)]^2 - 2 f(\mathbf{r}_1, \mathbf{r}_2) \right\} \rho_2^{HF}(\mathbf{r}_1, \mathbf{r}_2; \mathbf{r}_1, \mathbf{r}_2) \frac{1}{r_{12}} d\mathbf{r}_1 d\mathbf{r}_2 \qquad (58)$$

which, for the homogeneous electron gas, gives the following energy per electron:

$$\varepsilon_c = \int \left\{ [f(r_{12})]^2 - 2 f(r_{12}) \right\} \frac{1}{2} \rho g^{HF}(r_{12}) \frac{1}{r_{12}} d\mathbf{r}_{12} \qquad (59)$$

We have computed eq. (59) with two different values of $q$: the original one ($q = 2.29$), and the value proposed in the present paper ($q = q^*$). Incidentally, we note that the calculations have been done using either the HF pair-distribution or the gaussian ansatz: the differences in the results are smaller than 2% in the range of values of $r_s$ between 0.1 and 10 Bohrs.

As already reported in ref. [32], the original CS theory poorly accounts for the PW correlation energy. Using $q = q^*$ yields to a better approximation. However, the resulting curve overestimates the one obtained by the expression proposed in the present work, which, in turn, overestimates the PW one. For instance, at $r_s = 3$ Bohrs, our model allows to recover 78% of the PW correlation energy, vs. 67% for the CS theory (with $q = q^*$). This improvement may be attributed to the post IPA effects, which are partly taken into account in the present approach, unlike the original CS work.

Finally, let us notice that the fact that $q^*$ is found to be smaller than the original CS parameter shows that the Coulomb-hole is larger than predicted in the CS model, which artificially compensates for the neglected kinetic component of the correlation energy.

## IV Results for atoms and ions

Eq. (57) can be used as a local functional in DFT calculations. The corresponding correlation potential $v_c$, to be used in the Kohn and Sham equation, is derived from eqs. (57) and (53) by means of the relation [33]:

$$t_c = 3 v_c - 4 \varepsilon_c \qquad (60)$$

We have performed self-consistent non-spin-polarized calculations for the lightest atoms (till to $Z=18$) and for the rare gas atoms. The correlation energies that we have found are reported in table 2 and compared with results of similar calculations performed using the PW correlation functional [31]. In the same table are also reported the very accurate reference values obtained by Davidson and Chakravorty [34] for He to Ar atoms, supplemented by the values obtained by Clementi and Corongiu [35] for Kr and Xe. The latter are provided by the so-called "Virial constrained effective hamiltonian method", which is believed to give excellent estimates for correlation energies of atoms [35].

It is well known that the LDA overestimates the correlation energies of atoms by a factor of about 2. This is of course also the case of the PW expression. Not surprisingly Eq. (55) improves this situation: the discrepancies with respect to the reference values are considerably reduced, although they remain large. Furthermore, these discrepancies strongly decrease with the increase of the atomic number. The percentage error is 124% for He, approximately 50% for Ne and Ar, and reduces to 16% and 7% for Kr and Xe, respectively. It is interesting to notice that the opposite trend is found by the original CS theory [36]. We recall that CS optimized the $q$ parameter in order to obtain the exact total energy of He. Consequently, their approximation is very accurate for light atoms, and the errors do not exceed 4% and 2% for Ne and Ar, respectively. Increasing the atomic number, however, their approximation becomes progressively less accurate and the discrepancies with respect to the reference values increase (leading to percentage error of 16% for Kr and 20% for Xe, inasmuch as the estimated values for Kr and Xe can be considered as exact values). Finally, the improvements that we have found with respect to the PW results, are essentially the consequence of the fact that eq. (57) underestimates the absolute value of the correlation energy of the homogeneous electron gas. For this reason, they can not be considered as particularly significant.

It is much more interesting to look at quantities related to energy differences such as the ionization potentials. In fact, it is known that, even if the errors on the LDA total energies are quite large, the energy differences are, in many cases, very accurately obtained by this approximation.

The calculated ionization potentials obtained by the present approach and the PW implementation of the LDA are compared with the experimental data [37] in table 3. In order to make a consistent comparison, we have added to the ionization potentials, calculated by our method, the contributions due to the spin-polarization effects. The latter have been evaluated performing both spin-polarized and non-spin-polarized PW calculations. As we have anticipated, the PW-LDA results are quite accurate and their errors are generally smaller than 4% (with only two exceptions). Nevertheless, our approach improves these results: with the exception of Be and Al, the errors are smaller than 3% and, in several cases, they do not exceed 1%. The average percentage error is 1.6%, to be compared with 2.4% for the PW approximation. This improvement can be reasonably attributed to the fact that our approximation largely neglects the long-range correlation of the homogeneous electron gas. As the long-range correlation certainly affects atoms in a very different way than the homogenous gas, the present approach probably avoids the inclusion of some spurious effects in the description of such highly localized systems.

**Conclusions**

Aiming at partly correcting the drawbacks of the CS approximation extensively discussed in the literature, we have proposed a modified approach based on a correlated 1-RDM. Although approximate, the expression used for the 1-RDM includes effects beyond the independent pair approximation and leads to correctly normalized one-electron densities. Furthermore, the correlation kinetic energy, which is neglected in the original theory or possibly accounted for by an artificial underestimation of the electron-electron correlation potential energy, can be now coherently introduced in the theory. The application of the proposed approximated 1-RDM to the case of the homogeneous gas has allowed the derivation of an explicit expression of the correlation kinetic energy. The parameter $q$ entering in this expression has been obtained analytically by the comparison with the quasi-exact PW results and the requirement of having the correct sign in the limit $r_s \to 0$. Inverting the usual procedure, we have derived from the correlation kinetic energy the corresponding total correlation expression. When compared with the PW results, both the correlation kinetic and the total correlation energies show a satisfactory agreement at intermediate densities, but fail in the two limiting cases, low and high densities. This drawback in the case of the homogeneous gas results to be an advantage when the new expressions are used as a modified-local-density approximation in atomic calculations: both the correlation energies and the ionization potentials are strongly improved with respect to the corresponding results obtained by the standard LDA approximation. This suggests that the proposed functional could be also useful in molecular and solid-state calculations.

**Appendix**

A closed-form expression of the 1-RDM can not be deduced from the wave function given in eq. (21) owing to the presence of the factor $\prod_{i<j}[1-f(\mathbf{r}_i,\mathbf{r}_j)]$. Instead, one can consider the related pair-wise correlated wave function:

$$\psi(x_1,...,x_N) = \psi^{HF}(x_1,...,x_N)\left[1-\sum_{i<j} f(\mathbf{r}_i,\mathbf{r}_j)\right] \qquad (A1)$$

From this wave function, one can derive the following closed-form expression of the 1-RDM:

$$\begin{aligned}
\rho_1(\mathbf{r}_1;\mathbf{r}_1') &= \rho_1^{HF}(\mathbf{r}_1;\mathbf{r}_1') \\
&+ 2\int\left[-f(\mathbf{r}_1,\mathbf{r}_2)-f(\mathbf{r}_1',\mathbf{r}_2)+f(\mathbf{r}_1,\mathbf{r}_2)f(\mathbf{r}_1',\mathbf{r}_2)\right]\rho_2^{HF}(\mathbf{r}_1,\mathbf{r}_2;\mathbf{r}_1',\mathbf{r}_2)d\mathbf{r}_2 \\
&+ 3\int\{-2f(\mathbf{r}_2,\mathbf{r}_3)+2f(\mathbf{r}_1,\mathbf{r}_2)f(\mathbf{r}_1',\mathbf{r}_3)+2f(\mathbf{r}_1,\mathbf{r}_2)f(\mathbf{r}_2,\mathbf{r}_3)+2f(\mathbf{r}_2,\mathbf{r}_3)f(\mathbf{r}_1',\mathbf{r}_2) \\
&\qquad\qquad + [f(\mathbf{r}_2,\mathbf{r}_3)]^2\}\rho_3^{HF}(\mathbf{r}_1,\mathbf{r}_2,\mathbf{r}_3;\mathbf{r}_1',\mathbf{r}_2,\mathbf{r}_3)d\mathbf{r}_2 d\mathbf{r}_3 \\
&+ 12\int[f(\mathbf{r}_1,\mathbf{r}_2)f(\mathbf{r}_3,\mathbf{r}_4)+f(\mathbf{r}_1',\mathbf{r}_2)f(\mathbf{r}_3,\mathbf{r}_4) \\
&\qquad\qquad + 2f(\mathbf{r}_2,\mathbf{r}_3)f(\mathbf{r}_2,\mathbf{r}_4)]\rho_4^{HF}(\mathbf{r}_1,\mathbf{r}_2,\mathbf{r}_3,\mathbf{r}_4;\mathbf{r}_1',\mathbf{r}_2,\mathbf{r}_3,\mathbf{r}_4)d\mathbf{r}_2 d\mathbf{r}_3 d\mathbf{r}_4 \\
&+ 30\int f(\mathbf{r}_2,\mathbf{r}_3)f(\mathbf{r}_4,\mathbf{r}_5)\rho_5^{HF}(\mathbf{r}_1,\mathbf{r}_2,\mathbf{r}_3,\mathbf{r}_4,\mathbf{r}_5;\mathbf{r}_1',\mathbf{r}_2,\mathbf{r}_3,\mathbf{r}_4,\mathbf{r}_5)d\mathbf{r}_2 d\mathbf{r}_3 d\mathbf{r}_4 d\mathbf{r}_5 \qquad (A2)
\end{aligned}$$

Retaining just the two first terms in the right hand member of this equation, and introducing a normalization factor, one recovers eq. (26). The neglected terms involve an increasing number of electrons. It is obvious, from their analytic expressions, that they can be considered negligible excepted at high densities.

# References

[1] R. Colle and O. Salvetti, Theoret. Chim. Acta **37**, 329 (1975).

[2] This approach is sometimes referred to as the "correlation factor approach". See, for instance, F. Moscardó and A. J. Pérez-Jiménez, Int. J. Quantum Chem. **61,** 313 (1997).

[3] O.A.V Amaral and R. Mc Weeny, Theor. Chim. Acta **64**, 171 (1983).

[4] For an extensive list of references related to the Colle-Salvetti approach, see S. Caratzoulas, Phys. Rev. A **63**, 062506 (2001).

[5] R. Singh, L. Massa and V. Sahni, Phys. Rev. A **60**, 4135 (1999).

[6] S. Caratzoulas and P. J. Knowles, Mol. Phys. **98**, 1811 (2000).

[7] Some of these shortcomings can be corrected through simple ansatzs: see for instance the gaussian resummation ansatz proposed in ref. 4.

[8] H. Meyer, T. Müller, and A. Schweig, J. Mol. Struct. (Theochem) **360**, 55 (1996).

[9] F. A. Stevens, Jr. and M. A. Pokrant, Phys. Rev. A **8**, 990 (1973).

[10] A. Görling, M. Levy, and J.P. Perdew, Phys. Rev. B **47**, 1167 (1993).

[11] The normalization choice of P. O. Löwdin is adopted: Phys. Rev. **97**, 1474 (1955).

[12] For general properties of reduced density matrices, see E. R. Davidson, *Reduced Density Matrices in Quantum Chemistry*, Academic Press (1976).

[13] See the road map proposed by A. J. Thakkar, A. C. Tanner, and V. H. Smith, Jr. in: *Density Matrices and Density Functionals: Inter-relationships Between Various Representations of One-matrices and Related Densities: A Road Map and an Example*, edited by R.M. Erdahl and V.H. Smith, Jr. (Reidel, Dordrecht, Holland, 1987), pp. 327-337.

[14] This function is sometimes referred to as the reciprocal structure factor. See an example of analysis of such a function in: P. Patisson, W. Weyrich, J. Phys. Chem. Sol. **40**, 40 (1979).

[15] See also S. Ragot, J. M. Gillet, P. J. Becker, Phys. Rev. B **65**, 235115 (2002).

[16] P. Gori-Giorgi and P. Ziesche, ArXiv: cond-mat/0205342

[17] L. Cohen, C. Frishberg, C. Lee, and L. J. Massa, Int. J. Quantum Chem. Symp. **19**, 525 (1986).

[18] A. Soirat, M. Flocco and L. Massa, Int. J. Quantum Chem. **49**, 291 (1994).

[19] As discussed in ref. 12 pp. 117. This point is however discussed in terms of pair excitation operators instead of correlation functions.

[20] R. Morrison, Int. J. Quantum Chem. **46**, 583 (1993).

[21] See R. G. Parr and W. Yang, *Density-Functional Theory of Atoms and Molecules*, Oxford University Press (1989).

[22] P. Fulde, *Electron Correlations in Molecules and Solids*, Springer-Verlag (1993).

[23] C. Lee and R. G. Parr, Phys. Rev. A **35**, 2377 (1987).

[24] J. Meyer, J. Bartel, M. Brack, P. Quentin, and S. Aicher, Phys. Lett. B **172**, 122 (1986).

[25] M. Berkowitz, Chem. Phys. Lett. **129**, 486 (1986).

[26] Stephen Wolfram, Mathematica Version 4.0.

[27] M. Gell-Mann and K. A. Brueckner, Phys. Rev. **106**, 364 (1957).

[28] W. J. Carr Jr. and A. A. Maradudin, Phys. Rev. A **133**, 371 (1964).

[29] W. J. Carr, Jr, Phys. Rev. **122**, 1437 (1961).

[30] P. Nozieres and D. Pines, *The Theory of Quantum Liquids, Vol. I*, New York, Benjamin, (1966).

[31] J. P. Perdew and Y. Wang, Phys. Rev. B **45**, 13244 (1992), ibid. Phys. Rev. B **46**, 12947 (1992).

[32] J. Tao, P. Gori-Giorgi, J. P. Perdew and R. Mc Weeny, Phys. Rev. A **63**, 032513 (2001).

[33] See, for example, M. Levy and J. P. Perdew, Phys. Rev. A **32**, 2010 (1985).

[34] E. R. Davidson and S. J. Chakravorty. J. Phys. Chem. **100**, 6167 (1996).

[35] as published in: E. Clementi and D. W. M. Hofmann, THEOCHEM, 330, 17 (1995), see also, for a discussion of the method used and the reliability of the results obtained: E. Clementi and G. Corongiu. Int. Journal of Quant.Chem. 62, 571 (1996) and, for the "low-limit" MP2 estimates of the correlation energies: J. R. Flores, K. Jankowski and R. Slupski. Coll.of Czech. Chem. Com. 2003, 68, 240.

[36] C. Lee, W. Yang, and R. G. Parr, Phys. Rev. B **37**, 785 (1988).

[37] *Handbook of Chemistry and Physics* (1989-1990) 70[th] ed., R. C. Weast, Ed. (Chemical Rubber Company, Boca Raton, FL, 1990).

Table 1. Correlation kinetic energies per electron calculated by eq. (53) and by the Perdew and Wang expression.

| $r_s$ | PW | Present Work |
|---|---|---|
| 0.01 | 0.1595 | 0.0827 |
| 0.10 | 0.0918 | 0.0741 |
| 0.50 | 0.0511 | 0.0485 |
| 1.00 | 0.0367 | 0.0315 |
| 2.00 | 0.0246 | 0.0164 |
| 5.00 | 0.0124 | 0.0048 |
| 10.00 | 0.0066 | 0.0015 |
| 20.00 | 0.0032 | 0.0004 |

Table 2. Absolute values (in Hartrees) of the correlation energies of the lightest atoms and of the rare gas atoms. The reference values are the very accurate results reported in Ref. [34] completed, for Kr and Xe, by the results obtained by Clementi and Corongiu [35].

|    | Present work | PW    | Reference values |
|----|--------------|-------|------------------|
| He | 0.094        | 0.111 | 0.042            |
| Li | 0.130        | 0.150 | 0.045            |
| Be | 0.180        | 0.224 | 0.094            |
| B  | 0.237        | 0.289 | 0.12             |
| C  | 0.301        | 0.357 | 0.16             |
| N  | 0.370        | 0.426 | 0.19             |
| O  | 0.442        | 0.534 | 0.26             |
| F  | 0.517        | 0.638 | 0.32             |
| Ne | 0.594        | 0.740 | 0.39             |
| Na | 0.641        | 0.800 | 0.40             |
| Mg | 0.697        | 0.887 | 0.44             |
| Al | 0.753        | 0.962 | 0.47             |
| Si | 0.815        | 1.037 | 0.51             |
| P  | 0.881        | 1.112 | 0.54             |
| S  | 0.951        | 1.220 | 0.61             |
| Cl | 1.023        | 1.323 | 0.67             |
| Ar | 1.097        | 1.423 | 0.73             |
| Kr | 2.397        | 3.268 | 2.07             |
| Xe | 3.683        | 5.178 | 3.43             |

Table 3. Ionization potentials (in eV) of the lightest atoms and of the rare gas atoms. The experimental values are taken from Ref. [37]. The values in parenthesis are the percentage errors with respect to the experimental data.

|     | Present work |        | PW    |        | Exp.  |
|-----|--------------|--------|-------|--------|-------|
| Li  | 5.26         | (-2.4) | 5.47  | (1.5)  | 5.39  |
| Be  | 8.81         | (-5.5) | 9.03  | (-3.1) | 9.32  |
| B   | 8.36         | (0.7)  | 8.58  | (3.4)  | 8.30  |
| C   | 11.55        | (2.6)  | 11.76 | (4.4)  | 11.26 |
| N   | 14.77        | (1.7)  | 14.99 | (3.2)  | 14.53 |
| O   | 13.66        | (0.3)  | 13.90 | (2.1)  | 13.62 |
| F   | 17.81        | (2.2)  | 18.06 | (3.7)  | 17.42 |
| Ne  | 21.92        | (1.7)  | 22.18 | (2.9)  | 21.56 |
| Na  | 5.16         | (0.4)  | 5.37  | (4.5)  | 5.14  |
| Mg  | 7.51         | (-1.8) | 7.73  | (1.0)  | 7.65  |
| Al  | 5.79         | (-3.3) | 6.00  | (0.2)  | 5.99  |
| Si  | 8.06         | (-1.1) | 8.27  | (1.5)  | 8.15  |
| P   | 10.32        | (-1.6) | 10.53 | (0.4)  | 10.49 |
| S   | 10.33        | (-0.3) | 10.55 | (1.8)  | 10.36 |
| Cl  | 13.03        | (0.5)  | 13.25 | (2.2)  | 12.97 |
| Ar  | 15.71        | (-0.3) | 15.94 | (1.1)  | 15.76 |
| Kr  | 14.05        | (0.4)  | 14.27 | (1.9)  | 14.00 |
| Xe  | 12.35        | (1.8)  | 12.57 | (3.6)  | 12.13 |

# Figure captions

Figure 1. Comparison of the Hartree-Fock autocorrelation function $B^{HF}(s)$ and its gaussian approximation $B_G^{HF}(s)$.

Figure 2. Comparison of the correlation kinetic energy of the homogeneous electron gas calculated by various approximations.

Figure 3. Comparison of the total correlation energy of the homogeneous electron gas calculated by various approximations.

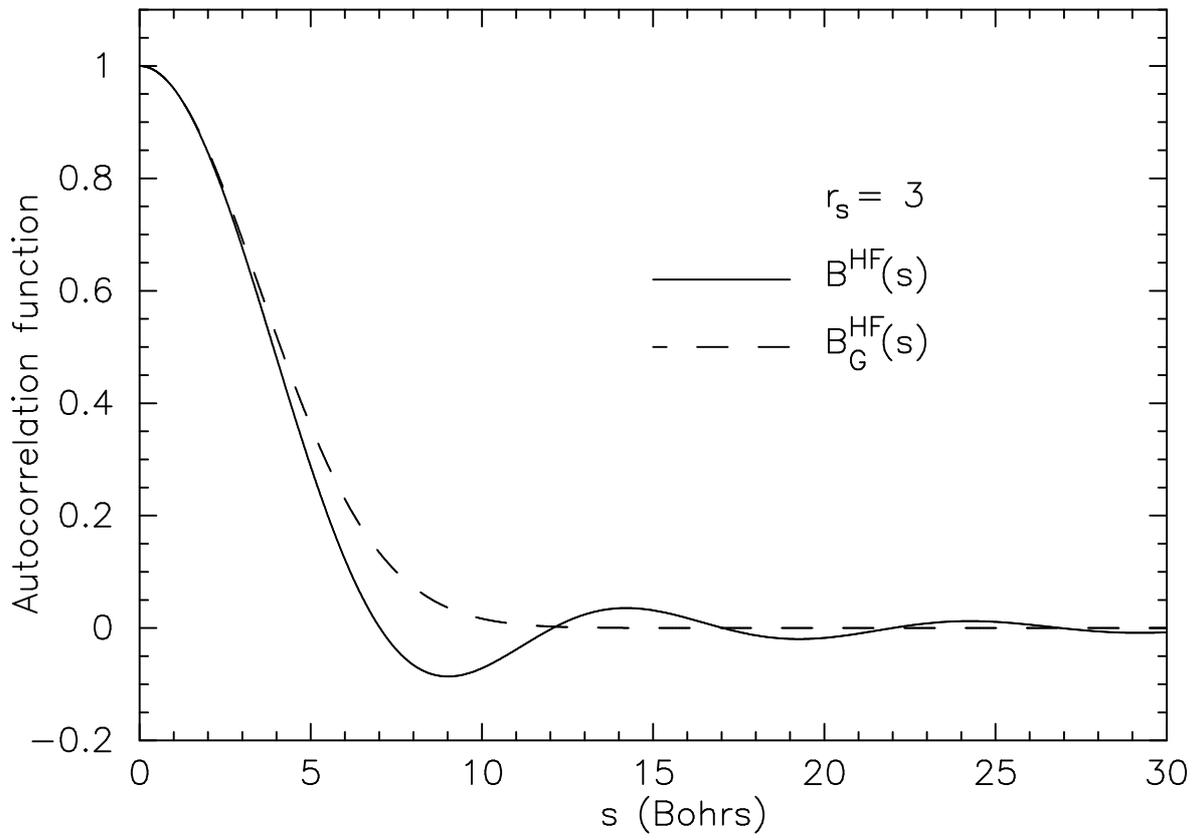

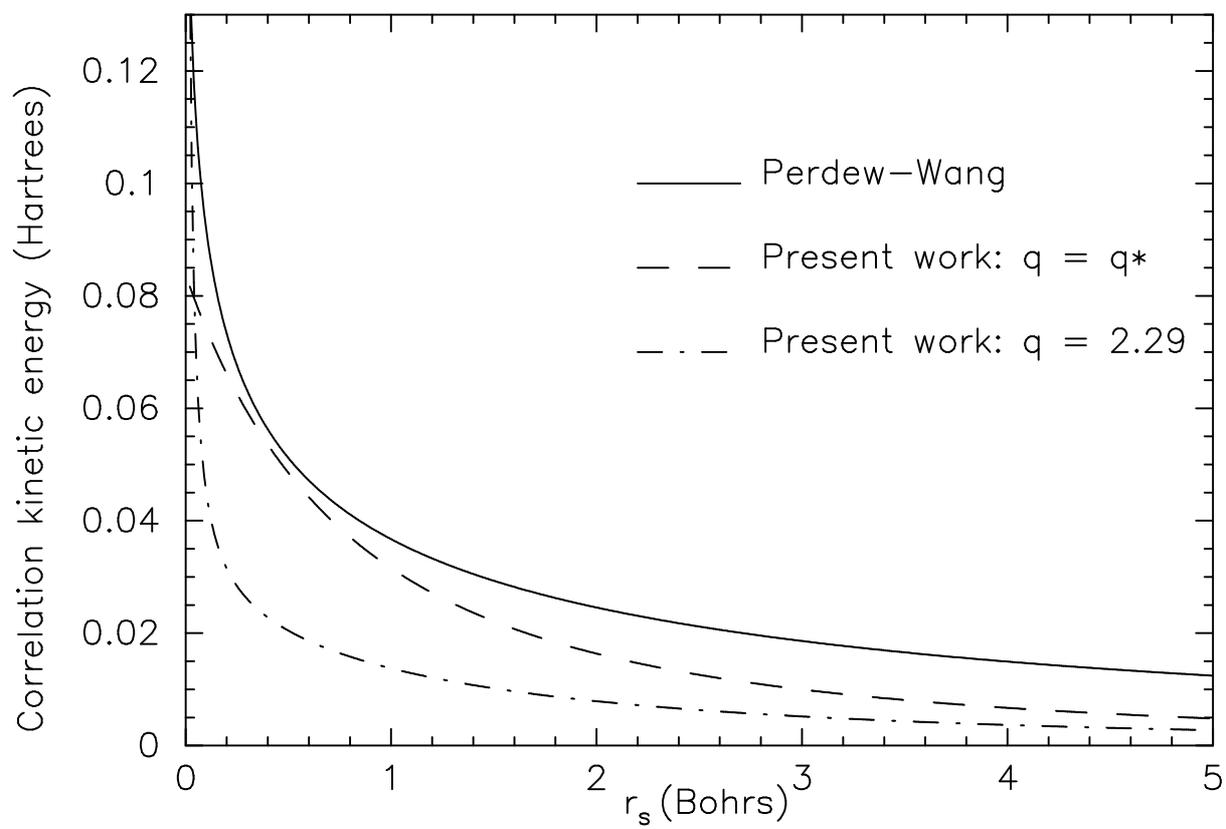

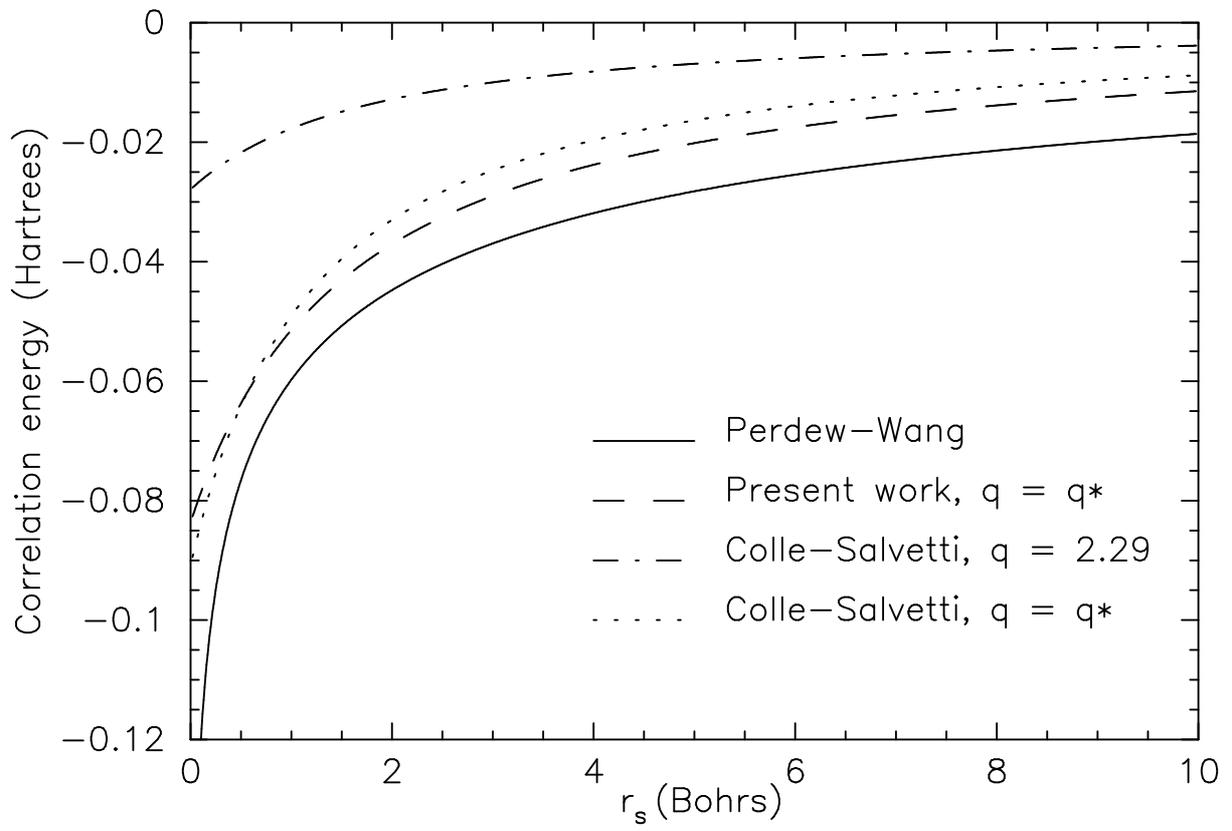